\documentclass{ws-ijmpa}
\usepackage[super,compress]{cite}
\usepackage{graphicx}
\usepackage{braket}
\usepackage{xcolor}
\usepackage[utf8]{inputenc}

\begin{document}

\title{Remarks on the preservation of no-signaling principle in parity-time-symmetric quantum mechanics}

\author{Bijan Bagchi}

\address{Department of Physics, Shiv Nadar University, Gautam Buddha Nagar\\
Dadri, Uttar Pradesh 201314, India\\
bbagchi123@gmail.com}

\author{Suvendu Barik}

\address{Department of Physics, Shiv Nadar University, Gautam Buddha Nagar\\
Dadri, Uttar Pradesh 201314, India\\
sb749@snu.edu.in}

\maketitle

\begin{history}
\received{Day Month Year}
\revised{Day Month Year}
\end{history}

\begin{abstract}
Working within the framework of parity-time-symmetric quantum mechanics we look into the possibility of entanglement generation and demonstrate that the feature of non-violation of no-signaling principle may hold for the simplest non-trivial case of bipartite systems.  Basically our arguments are based on the computation of the reduced density matrix of one party
to justify that the entropy of the other does not change.

\end{abstract}

\ccode{PACS numbers: 03.67.Bg, 03.65.Ud, 03.65.Ta}

\section{Introduction}
The notion of quantum entanglement \cite{NC, BZ, WGE, PKP} speaks of a shared existence of particles having their properties interlinked with each other. An interesting manifestation of entanglement is that the correlation survives even when the particles move to a large distance after having come into contact. Different aspects of quantum entanglement have been studied and a substantial literature has accumulated in this subject (see, for instance, \cite{SS, Wan}). In the $PT$-symmetry context an early attempt was made in \cite{Wan}. In this article we push the issue a little further by addressing quantum entanglement in the framework of a complex extension of quantum mechanics concentrating on the special class of complex parity ($P$)-time ($T$)-symmetric Hamiltonians. 
\par
Almost two decades ago, Bender and Boettcher\cite{BB1} proposed a special class of non-Hermitian Hamiltonians, which were manifestly 
($PT$)-symmetric, that support a real bound-state spectrum. The interplay between the parametric regions where $PT$ is unbroken and
the ones in which it is not, as signaled by the appearance of conjugate-complex eigenvalues, has also found experimental
support through the observations of a phase transition\footnote{The transition refers to the breaking of $PT$-symmetry when exceptional points appear.} that clearly marks out the separation of these regions (see, for example, \cite{Gan} and earlier references therein): in particular, balancing gain and loss of certain experimental properties has uncovered the relevance of $PT$-symmetric structures in such systems \cite{ZM1, ZM2, KM}.
\par
In standard quantum mechanics ($SQM$), a coherent description of physics is possible when Dirac's norm is employed \cite{Drc1}. Of course there could be other norms (see \cite{CMB} for their physical implications). However these norms run into one problem or another. In particular,  $PTQM$ systems are generally plagued with negative norms. The reason lies in the difference in the definition of the inner product in $SQM$ as introduced by Dirac namely,
\begin{equation}
(f,g) \equiv \int_\Re dx[Tf(x)]g(x),  \quad f,g \in L_2(\Re)
\end{equation}
where $Tf(x)=f^*(x)$, and that of $PTQM$ namely,
\begin{equation}
(f,g)_{PT} \equiv \int_\Re dx[PT f(x)]g(x),  \quad f,g \in L_2(\Re)
\end{equation}
where one defines $PTf(x)=[f(-x)^*]$. The above definition of $PT$-norm very often leads to an indefinite norm implying that a $PT$-systems lacks a viable probabilistic interpretation \cite{Bag1, Jap1}.
\par
However, an introduction of a linear operator $C$ to construct a $CPT$-inner product \cite{BB4} in the following sense
\begin{equation}
\label{eq:3} 
(f,g)_{CPT} \equiv \int_\Re dx[CPT f(x)]g(x)
\end{equation}
with the positive-definiteness of the associated norm,
enables one to get rid of this handicap. Note that $C$ commutes with both
the Hamiltonian and the operator $PT$. Further it is idempotent and has eigenvalues $\pm 1$.
A $PT$-symmetric system is supposed to evolve in a manner such that the accompanying time evolution of the state vector
is unitary with respect to the $CPT$ inner product. For a plausible construction of the $C$-operator see \cite{Bag2}.
\par
We propose that no-signaling principle holds for bipartite systems whenever one of the subsystems is $PT$-symmetric. In a different context, such a study\cite{Jap2} has led to the reproduction of the Clauser-Horne-Shimony-Holt (CHSH) inequality in connection with the invariance of the entanglement. Interestingly, an experimental search seems to put in evidence that a simulated $PT$ symmetric subsystem preserves no-signaling\cite{SB4}. However, theoretical results pointing to the contrary have also been noted\cite{SB1,SK1}. Here, we must emphasize that in the bipartite scenario, no-signaling means that, for two observers, say Alice and Bob, whatever Alice does the outcome probability of any measurement by Bob is unchanged. This is the central assumption\footnote{$PT$-symmetry is well established in the single-party case but for a multi-partite scenario the problem of finding entangled states is NP-hard\cite{WGE}. It insists on the resolution for bipartite systems first.} that is implicit in our paper. In $SQM$, where the concept of Hermiticity\footnote{Hermiticity is known by the condition satisfied by an operator associated with an observable.} holds confirming the reality of the associated energy spectrum, one shows that the outcome probability of any measurement by Bob is determined entirely by his reduced density matrix. Consequently, no signaling can be proven by showing that the reduced density matrix of Bob is
unchanged whatever operation Alice does.
\par
In the present work, we will consider a two-dimensional example of a $PT$-Hamiltonian and compute the reduced density matrix of one party using the definition of $CPT$-inner product as just given. We will then show that the entanglement entropy of the density matrix remains unaltered after applying a time evolution operator on them. It should be noted that in our work we assumed the case of the even time-reversal operator which is valid for bosonic systems \cite{REF1}.
\par
It needs to be mentioned that similar results as ours concerning the no-signaling principle were obtained by the authors of \cite{Jap2} , in which the former is shown for bipartite systems where either one or both Hamiltonians of the subsystems are non-Hermitian and $PT$-symmetric, as defined in a space of states controlled by a $CPT$ inner product. However the present work differs from their approach in our use of modified density matrices, leading to an appropriate reasoning for the estimation of the reduced density matrices of the parties by the employment of the $CPT$ inner product for the finite representation of the subsystems. We further highlight that the work of \cite{Jap2} suffered the lack of reasoning to establish the reduced density matrices calculated under the modified norm as an appropriate quantity in realising the local measurements made by the observer within the $PTQM$ framework. However, this work demonstrates to solve precisely that.

\section{Prerequisites}
\subsection{Finite representation of unbroken $PT$-symmetric systems}
For the operators in a $PTQM$ theory it needs to be noted that their eigenstates are not orthogonal under the standard Dirac inner product. An implication is that the Hamiltonian given by 
\begin{equation}
\label{rep}
    H = \sum_{i=1}^{N} \lambda_{i}\ket{\psi_{i}}\bra{\phi_{i}}
\end{equation}
with eigenvalues $\{\lambda_{i}|1\leq i \leq N\}$ and eigenstates $\{\ket{\psi_{i}}|1 \leq i \leq N\}$ along with $\sum_{i=1}^{N}\ket{\psi_{i}}\bra{\phi_{i}} = \mathbb{I}$, can
never coincide with the more familiar 

\begin{equation}
\label{incompref}
    H_{S} = \sum_{i=1}^{N} \lambda_{i}\ket{\psi_{i}}\bra{\psi_{i}}
\end{equation}
which satisfies the Hermiticity condition of $SQM$.
\par
The utmost  we can do is to project $H$ in a factorized form involving $H_S$ as one of the factors 
\begin{equation}
\label{repnewform}
    H = \left(\sum_{i=1}^{N} \lambda_{i}\ket{\psi_{i}}\bra{\psi_{i}}\right)\left(\sum_{j=1}^{N} \ket{\phi_{j}}\bra{\phi_{j}}\right) = H_{S}\hat{\eta}
\end{equation}
where $\hat{\eta}$ denoting the sum
\begin{equation}
\hat{\eta} =\sum_{j=1}^{N} \ket{\phi_{j}}\bra{\phi_{j}}.
\end{equation}
The feature of $H$ is that it commutes with the $PT$ operator defined by
\begin{equation}
    PT = \sum_{i=1}^{N} \alpha_{i}\ket{\psi_{i}}\bra{\phi_{i}}.
\end{equation}
Consider now an operator $A$ admitting factorization
\begin{equation}
    A = A_{S}\hat{\eta},\;\;\;\;A_{S}=\sum_{i=1}^{N}\lambda_{i}^{'}\ket{n_{i}}\bra{n_{i}},\;\;\;\;\ket{n_{i}}=\sum_{j=1}^{N}c_{ij}\ket{\psi_{j}}
\end{equation}
where $\lambda_{i}^{'}$'s are the coefficient constants in the expansion. We can then write corresponding to the modified inner product
\begin{equation}
\label{expect1}
    (\psi,A\psi)_{\hat{\eta}} = \braket{\psi|\hat{\eta} A_{S}\hat{\eta}|\psi} = \braket{\phi|A_{S}|\phi}.
\end{equation}
This makes the role of $\hat{\eta}$ clear. With $H$ defining the Hamiltonian of the system, any valid measurement corresponding to the operator $A$ over the energy eigenstates satisfies the pseudo-hermiticity relationship\cite{AM1, AM2}
\begin{equation}
\label{pseher}
   \hat{\eta}H\hat{\eta}^{-1} = H^{\dagger}.
\end{equation}
The expectation value of $A$ over the state $\psi$ in a pseudo-Hermitian framework\cite{AM2} is the same as the expectation value of $A_{S}$ over the state $\phi$ in the standard framework. 
\par
As in $SQM$, the outcome of measurements performed corresponding to the operator $A_{S}$ solely depend on the state $\ket{\phi}$ and given by the density matrix $\rho = \ket{\phi}\bra{\phi}$ namely,
\begin{equation}
   \braket{\phi|A_{S}|\phi} = \text{Tr}\left(\ket{\phi}\bra{\phi}A_{S}\right).
\end{equation}
Using the cyclic property of the trace, $\text{Tr}\left(\ket{\phi}\bra{\phi}A_{S}\right) = \text{Tr}\left(\hat{\eta}\ket{\psi}\bra{\phi}A_{S}\right) = \text{Tr}\left(\ket{\psi}\bra{\phi}A_{S}\hat{\eta}\right) = \text{Tr}\left(\ket{\psi}\bra{\phi}A\right)$, and also \eqref{expect1}, we can express the $\eta$-inner product to be 
\begin{equation}
 (\psi,A\psi)_{\hat{\eta}} = \text{Tr}\left(\ket{\psi}\bra{\phi}A\right).
\end{equation}
The effective density operator of $\ket{\psi}$ in a pseudo-hermitian framework is thus $\ket{\psi}\bra{\phi} = \ket{\psi}\bra{\psi}\hat{\eta}$. Note that when $\hat{\eta}=\mathbb{I}$ we recover the standard result of $SQM$.   
\par
We now provide a scheme to calculate overlaps under a $CPT$-inner product by introducing a generalized inner-product  restricting to finite Hilbert space. For any operator $\hat{X}$ which shares simultaneous eigenstates with the Hamiltonian $\hat{H}$, the $X$-inner product is defined to be $\braket{-|-}_{X}$ and obeys
\begin{equation}
\label{eq:Xnorm}
    \braket{\phi|\psi}_{X} = \hat{X}\ket{\phi}\cdot\ket{\psi}, \quad \bra{\phi}_{X} = (\hat{X}\ket{\phi})^{T}.
\end{equation}
Replacing $\hat{X}$ by the $CPT$-operator, the above equation showcases an appropriate way to perform calculations when a  $CPT$ inner product is invoked. Indeed, one could derive
\begin{equation}
   \braket{\phi|\psi}_{CPT} = CPT\ket{\phi}\cdot\ket{\psi} = \braket{\phi|(CP)^{T}|\psi}.
\end{equation}
which implies that the intertwining operator $\hat{\eta}$ for an unbroken $PT$-symmetric system is $(CP)^{T}$. For the density operator $\rho$ in the state $\ket{\psi}$ we deduce easily that
\begin{equation}
  \rho = \ket{\psi}\bra{\psi}_{CPT} = \ket{\psi}\bra{\psi}(CP)^{T}.  
\end{equation}
With the above results at hand, we analyze in the next section the no-signaling principle in a $PT$-symmetric framework.

\subsection{Entanglement and the no-signaling principle in a $PT$-symmetric framework}
For $N$ quantum subsystems defined over a set of Hilbert spaces $\{\mathcal{H}_{i}|1\leq i\leq N\}$, a composite system generated out of these subsystems will exist in $\mathcal{H}_{1}\otimes\mathcal{H}_{2}\otimes...\otimes\mathcal{H}_{N}$. Let us define a joint state as the tensor product 
\begin{equation}
 \ket{\psi} = \ket{\psi_{1}}\otimes\ket{\psi_{2}}\otimes...\otimes\ket{\psi_{N}}\in \mathcal{H},\;\;\;\;\ket{\psi_{i}}\in\mathcal{H}_{i}.
\end{equation}
A pure state $\ket{\psi}$ of $\mathcal{H}$ such as the one given above is said to be separable \cite{Sib}. A state of $\mathcal{H}$ which is non-separable is called an entangled pure state\footnote{This criterion also holds for an infinite representation of subsystems.}. 
\par
For $N=2$, which conforms to a bipartite system, a measure of entanglement is provided by the following definition of information entropy
\begin{equation}
E(\psi) = -\mathrm{Tr}_{1}(\rho_{1}\log_{2}\rho_{1}) = -\mathrm{Tr}_{2}(\rho_{2} \log_{2}\rho_{2})
\end{equation}
where $\rho$ is the density matrix corresponding to $\ket{\psi}$ and the reduced density matrices $\rho_{1}$ and $\rho_{2}$ are given  respectively by the partial traces of $\rho$:  $\rho_{1} = \mathrm{Tr}_{2}(\rho)$ and $\rho_{2} = \mathrm{Tr}_{1}(\rho)$. The entropy $E(\psi)$ is
\begin{equation}
E(\psi) =  -\sum_{i} \lambda_{i}\log_{2}{\lambda_{i}}    
\end{equation}
where $\lambda_{i}$'s are the eigenvalues of the relevant reduced density matrix. The scheme of calculating the density matrix of the states of the system will, however, vary if we dealing with pseudo-hermitian subsystems.\footnote{One can equivalently perform the calculation of the density operator by following the scheme for the bra vector as provided in \eqref{eq:Xnorm}}
\par
Consider $\{\ket{u_{n}}\}$ and $\{\ket{v_{n}}\}$ as basis sets of the respective Hilbert spaces $\mathcal{H}_{1}$ and $\mathcal{H}_{2}$. The basis set of the composite Hilbert space $\mathcal{H}_{1}\otimes\mathcal{H}_{2}$ is then $\{\ket{u_{n}}\otimes\ket{v_{n}}\}$. As such, an entangled of a pure bipartite state is given by
\begin{equation}
\label{eq:purebipartite}
\ket{\psi} = \sum_{n,m=1}^{D_{1},D_{2}} C_{nm}\ket{u_{n}}\otimes\ket{v_{m}}, \quad \sum_{n,m} \mid C_{nm}\mid^{2} = 1   
\end{equation}
where $D_{1},D_{2}$ are the respective dimensions of the Hilbert spaces and $C_{nm}$ are constants. Since we restrict to bipartite systems only we take in what follows,  $D_{1}=D_{2}=2$.

\subsection{Overview of a $2\times2$ $PT$-Symmetric model}
For calculational simplicity we adopt the following form\footnote{This structure is equivalent to the matrix considered in \cite{BB4} modulo an identity factor.} of a two-level $PT$-symmetric Hamiltonian\cite{Jog}
\begin{equation}
\label{eq:Hamiltonian}
\hat{H}=\begin{pmatrix} 
i\gamma & -\zeta\\          
-\zeta & -i\gamma
\end{pmatrix}
\end{equation}
where $\gamma >0$ and $\zeta >0$.  With representation of the parity operator being $\hat{P} = \begin{bmatrix} 0&1 \\ 1&0 \end{bmatrix}$ and the time-reversal operation transforming like $T:i \rightarrow -i$, the $PT$-symmetric character of $\hat{H} $ is evident. 
\par
Because of the underlying $PT$-symmetry, the right and left eigenvectors of $\hat{H}$ are not the same. Specifically, the right eigenvectors (for $\left|\frac{\gamma}{\zeta}\right|\leq1$) read 
\begin{equation}
\ket{\psi_{\pm}} = \frac{1}{\sqrt{2\cos{\phi}}}
\begin{pmatrix}1 \\ \mp e^{\mp i\phi}\end{pmatrix}
\end{equation}
\\
where $\sin \phi = \frac{\gamma}{\zeta}$ and the eigenvalues of $\hat{H}$ are
\begin{equation}
\lambda_{\pm} = \pm\sqrt{\zeta^2-\gamma^2}
\end{equation}
These are entirely real if the inequality $\gamma < \zeta$ holds. The degeneracy of the eigenvalues takes place when $\gamma = \zeta$. However for $\gamma > \zeta$ the eigenvalues become purely imaginary complex conjugates.
\par
Following \cite{BB4}, we adopt, up to a sign, the $\hat{C}$ operator in the form
\begin{equation}
\hat{C} = \begin{bmatrix}-i\tan{\phi}&\sec{\phi}\\\sec{\phi}&i\tan{\phi}\end{bmatrix}.    
\end{equation}
\\
It is immediately verified that the actions of $\hat{P}\hat{T}$ and $\hat{C}$ operators on the eigenstates $\ket{\psi_{\pm}}$ work as
\begin{equation}
\begin{split}
    \hat{P}T&\ket{\psi_{\pm}} = \frac{\mp e^{\pm i\phi}}{\sqrt{2\cos{\phi}}}\begin{bmatrix}1 \\ \mp e^{\mp i\phi}\end{bmatrix},  \\
  \hat{C}&\ket{\psi_{\pm}} = \frac{\mp1}{\sqrt{2\cos{\phi}}}\begin{bmatrix}1 \\ \mp e^{\mp i\phi}\end{bmatrix}.
\end{split}    
\end{equation}
\\
These lead to the positive definiteness of the $CPT$-inner product for an arbitrary state $\ket{\psi} = \begin{bmatrix}a\\b\end{bmatrix} = \begin{bmatrix}r_{a}e^{i\theta_{a}}\\r_{b}e^{i\theta_{b}}\end{bmatrix}$ which is given using
\begin{equation}
\hat{C}\hat{P}T\ket{\psi} = \frac{1}{\cos{\phi}}\begin{bmatrix}a^{*}-i b^{*}\sin{\phi}\\b^{*}+i a^{*}\sin{\phi}\end{bmatrix} \\ 
\end{equation}
as
\begin{equation}
\begin{split}
\braket{\psi|\psi}_{CPT} &= \frac{1}{\cos{\phi}}[aa^{*}+bb^{*} - i(b^{*}a-a^{*}b)\sin{\phi}] \\&= \frac{1}{\cos{\phi}}[r_{a}^{2}+r_{b}^{2}+2r_{a}r_{b}\sin{\phi}\sin{(\theta_{a}-\theta_{b})}]\geq 0,
\end{split}
\end{equation}
consistent with the result obtained in \cite{SB3}.
\par
Finally, we might keep in mind that as $\phi\rightarrow 0$, the framework of $PT$-symmetric quantum mechanics ($PTQM$) transits to that of ($SQM$):
\begin{equation}
\label{PTQMQM}
    \begin{split}
    & \hat{H}\rightarrow-\zeta\sigma_{x}
     \\& \frac{1}{\sqrt{2\cos{\phi}}}
\begin{pmatrix}1 \\ \mp e^{\mp i\phi}\end{pmatrix} \rightarrow  \frac{1}{\sqrt{2}}
\begin{pmatrix}1 \\ \mp 1\end{pmatrix} 
    \\& \pm\sqrt{\zeta^2-\gamma^2} \rightarrow \pm\zeta
    \\& \hat{C} = \begin{bmatrix}-i\tan{\phi}&\sec{\phi}\\\sec{\phi}&i\tan{\phi}\end{bmatrix} \rightarrow \begin{bmatrix}0&1\\1&0\end{bmatrix} = \hat{P}
    \\& \braket{-|-}_{CPT} \rightarrow \braket{-|-}_{T}
    \end{split}
\end{equation}
where we identify $\braket{-|-}_{T}$ as the usual Dirac norm. It is useful to note that the Hamiltonian $\hat{H}$ also affords the following factorized representation
\begin{equation}
\hat{H} = \hat{H}_{QM}\hat{\eta},\;\;\;
\hat{H}_{QM} = -\zeta\cos{\phi}\begin{bmatrix}0&1\\1&0\end{bmatrix},\;\;\;\hat{\eta}=\begin{bmatrix}\sec{\phi}&i\tan{\phi}\\-i\tan{\phi}&\sec{\phi}\end{bmatrix}=(CP)^{T}.  
\end{equation}

\section{Entanglement in $PT$-symmetric systems}
In $SQM$, as the density operator of a bipartite system evolves,  no-signaling is established by showing the invariance of entropy for the initial and final entangled state. For an unbroken $PT$-symmetric system, however, the same can be established by adopting states to conform to pseudo-hermitian transformations. Specifically, the following steps are followed:
\begin{enumerate}
    \item First, for a pure entangled state $\ket{\psi}$, we determine the initial estimate of the quantity $E(\psi_{t=0})$.
    \item Then, we apply the time evolution operator on the composite state in the given Hilbert space $\mathcal{H}$ and calculate the reduced density matrix by performing partial traces of the density operator. It puts us in a position to determine the entanglement measure of $\psi_{t}$.
    \item Finally, we determine the time-dependent quantity $E(\psi_{t})$ to demonstrate the invariant result $E(\psi_{t=0}) = E(\psi_{t=t'})$.
\end{enumerate}
We now proceed to address the different subsystems as alluded to above.

\subsection{Subsystems governed by $PTQM$}
We focus on two subsystems each controlled by $PTQM$ according to
\begin{equation}
\label{eq:Hamiltonians}
\hat{H}_{1} = \begin{bmatrix}i\gamma&-\zeta \\ -\zeta&-i\gamma\end{bmatrix},  \quad \hat{H}_{2} = \begin{bmatrix}i\gamma'&-\zeta' \\ -\zeta'&-i\gamma'\end{bmatrix}
\\[0.5ex]
\end{equation}
with one Hamiltonian for each subsystem. The associated time evolution operator $U(t)=e^{-i\hat{H_i}t}, i=1,2$ maps $\hat{H}_{1}$ and $\hat{H}_{2}$ to their time-dependent forms. The eigenstates (normalised under $CPT$-inner product) of $\hat{H}_{i}$, which serve as a basis set of $\mathcal{H}_{i}$, $i=1,2$, are given by 
\begin{equation}
\begin{split}
\{\ket{u_{1}},\ket{u_{2}}\}&= \left\{\frac{1}{\sqrt{2\cos{\phi}}}\begin{bmatrix}1\\- e^{- i \phi}\end{bmatrix},\frac{1}{\sqrt{2\cos{\phi}}}\begin{bmatrix}1\\+ e^{+ i \phi}\end{bmatrix}\right\}
\\[0.75ex]\{\ket{v_{1}},\ket{v_{2}}\}&=  \left\{\frac{1}{\sqrt{2\cos{\phi'}}}\begin{bmatrix}1\\- e^{- i \phi'}\end{bmatrix},\frac{1}{\sqrt{2\cos{\phi'}}}\begin{bmatrix}1\\+ e^{+ i \phi'}\end{bmatrix}\right\}
\end{split}
\end{equation}
where $\sin{\phi} = \frac{\gamma}{\zeta}$ and $\sin{\phi'} = \frac{\gamma'}{\zeta'}$. We now construct an entangled state and look into its behaviour upon the application of the time evolution operator $\mathbb{I}\otimes U(t)$ where for concreteness we take $U(t) = e^{-i\hat{H_{2}}t}$. To this end, using the definition of \eqref{eq:purebipartite} we first arrive at the form
\begin{equation}
\label{eq:entstate}
\ket{\psi} = \sum_{n,m=1}^{2,2} C_{nm}\ket{u_{n}}\otimes\ket{v_{m}}, \quad \sum_{n,m} \mid C_{nm}\mid^{2} = 1.   
\end{equation}
having its bra counterpart reading 
\begin{equation}
\bra{\psi}=    \left(\hat{C}\hat{P}T\otimes\hat{C}\hat{P}T\ket{\psi}\right)^{T} = \sum_{n,m=1}^{2,2} C_{nm}^{*}\bra{u_{n}}_{CPT}\otimes\bra{v_{m}}_{CPT}.
\end{equation}
with $X$ replaced by $CPT$ in the scheme formulated in \eqref{eq:Xnorm}. Although of no direct concern here, the above notation of bra would be useful to handle entanglement for the multi-partite cases where the individual subsystems contribute towards defining an overall inner product of the composite system.
\par
The full density matrix of the entangled state which reads
\begin{equation}
\label{eq:denfull}
\rho_{1,2} = \sum_{n,m,a,b=1}^{2} C_{ab}C_{nm}^{*}\ket{u_{a}}\bra{u_{n}}_{CPT}\otimes\ket{v_{b}}\bra{v_{m}}_{CPT}.    
\end{equation}
has the individual elements as summarized below
\begin{equation}
\begin{split}
\braket{u_{i}|u_{j}}_{CPT} = \delta_{ij}&\\[0.5ex]\ket{u_{1}}\bra{u_{1}}_{CPT} = \frac{1}{2\cos{\phi}}\begin{bmatrix}e^{i\phi}&-1\\-1&e^{-i\phi}\end{bmatrix}&\\[0.5ex] \ket{u_{2}}\bra{u_{2}}_{CPT} = \frac{1}{2\cos{\phi}}\begin{bmatrix}e^{-i\phi}&1\\1&e^{i\phi}\end{bmatrix}&\\[0.5ex]\ket{u_{1}}\bra{u_{2}}_{CPT}= \frac{1}{2\cos{\phi}}\begin{bmatrix}e^{-i\phi}&1\\-e^{-2i\phi}&-e^{-i\phi}\end{bmatrix}&\\[0.5ex]\ket{u_{2}}\bra{u_{1}}_{CPT} = \frac{1}{2\cos{\phi}}\begin{bmatrix}e^{i\phi}&-1\\e^{2i\phi}&-e^{i\phi}\end{bmatrix}&. \end{split}  \end{equation}
These correspond to $\mathcal{H}_{1}$. A similar set can be found for $\mathcal{H}_{2}$ by replacing $u$ by $v$ and $\phi$ by $\phi'$. Concerning the trace of density operators it suffices to mention that it follows the usual results of normalized eigenstates for an appropriate inner product. 
\par
Applying the partial trace in $\mathcal{H}_{2}$ gives us the reduced density operator for $\mathcal{H}_{1}$
\begin{equation}
\rho_{1} = \mathrm{Tr}_{2}[\rho_{1,2}] = \sum_{a,b,n=1}^{2} C_{ab}C_{nb}^{*}\ket{u_{a}}\bra{u_{n}}_{CPT}
\end{equation}
where $\rho_{1}$ stands for the matrix
\begin{equation}
\label{eq:den1}
\begin{split}
&\frac{1}{2\cos{\phi}}\begin{bmatrix}
N_{11}& N_{12}\\N_{21}&N_{22}
\end{bmatrix} 
\end{split}
\end{equation}
whose elements read explicitly
\begin{equation}
\label{eq:denterm}
\begin{split}
    & N_{11} = (\alpha+\gamma)e^{i\phi} + (\beta+\delta)e^{-i\phi},\\
    & N_{12} = (\beta+\delta-\alpha-\gamma),\\
    & N_{21} = (\delta-\alpha)-\beta e^{-2i\phi} + \gamma   e^{2i\phi},\\
    & N_{22} = (\delta-\gamma)e^{i\phi}+(\alpha-\beta)e^{-i\phi},\\ 
    & \alpha = C_{11}C_{11}^{*} + C_{12}C_{12}^{*},\\
    & \beta = C_{11}C_{21}^{*} + C_{12}C_{22}^{*},\\
    & \gamma = C_{21}C_{11}^{*} + C_{22}C_{12}^{*} = \beta^{*},\\
    & \delta = C_{21}C_{21}^{*} + C_{22}C_{22}^{*} \;\;\;\;\text{and}\\
    & \alpha+\delta=1.
\end{split}    
\end{equation}
What happens when we apply\footnote{The act of operation of time evolution is equivalent to making a measurement on the entangled state by the first party here.} the time evolution operator on the density matrix of \eqref{eq:entstate}? A straightforward calculation gives   
\begin{equation}
\begin{split}
&\ket{\psi_{t}} = \mathbb{I}\otimes e^{-i\hat{H}_{1}t}\ket{\psi} = \sum_{n,m=1}^{2,2} e^{-i\lambda_{m}t}C_{nm}\ket{u_{n}}\otimes\ket{v_{m}}, \\ & \lambda_{1} = \sqrt{\zeta'^2-\gamma'^2},\;\;\;\lambda_{2} = -\sqrt{\zeta'^2-\gamma'^2}
\end{split}
\end{equation}
resulting in the following time-dependent form $\rho_{1,2}$
\begin{equation}
\rho_{1,2}(t) = \sum_{n,m,a,b=1}^{2}e^{i(\lambda_{m}-\lambda_{b})t} C_{ab}C_{nm}^{*}\ket{u_{a}}\bra{u_{n}}_{CPT}\otimes\ket{v_{b}}\bra{v_{m}}_{CPT}.    
\end{equation}
In particular $\rho_{1}(t)$ is expressible as
\begin{equation}
\label{eq:den2}
\rho_{1}(t) = \frac{1}{2\cos{\phi}}\begin{bmatrix}
N_{11}&N_{12} \\ N_{21}&N_{22}
\end{bmatrix} 
\end{equation}
following the convention set up in \eqref{eq:denterm}.
\par
We therefore obtain the result that \eqref{eq:den1} and \eqref{eq:den2} are the reduced density matrices corresponding to $\mathcal{H}_{1}$ respectively holding before and after the operation of time evolution operator. It shows invariance of the measurement made by the system guided by $\mathcal{H}_{2}$ i.e. $E(\psi)=E(\psi_{t})$. Thus no-signaling is a valid criterion in $PTQM$.
\par
To inquire as to whether the eigenvalues of the reduced density operators undergo any change if we transform to the standard $QM$ formalism, the answer is self-explanatory if we look at the dependence of the eigenvalues ($\omega_{\pm}$) on the parameters of the Hamiltonian \eqref{eq:Hamiltonians}. It is easily seen that
\begin{equation}
\label{eq:eigPT}
\begin{split}
\omega_{\pm} &= \frac{1}{2}\left((\alpha+\delta)\pm\sqrt{1+4(\beta\gamma-\alpha\delta)}\right)
\\&= \frac{1}{2}\left(1\pm\sqrt{1-4\mid C_{11}C_{22}-C_{12}C_{21}\mid^{2}}\right)
\end{split}
\end{equation}
implying that the parameters stay invariant.

\subsection{Subsystems governed by $PTQM$ and $SQM$}
For concreteness let the Hamiltonian $\mathcal{H}_{1}$ be relevant for the $PTQM$ while $\mathcal{H}_{2}$ holds for the $SQM$ system. For the latter we choose it to be represented by $\sigma_{x}$ whose eigenstates act as the basis states,
under the usual inner product definition, are
\begin{equation}
\{\ket{v_{1}},\ket{v_{2}}\} =  \{\ket{1},\ket{0}\} = \left\{ \frac{1}{\sqrt{2}}\begin{bmatrix}1\\-1\end{bmatrix},\frac{1}{\sqrt{2}}\begin{bmatrix}1\\1\end{bmatrix}\right\}.     
\end{equation}
The inner product structure, which is the same as in $SQM$, shows 
\begin{equation}
\begin{split}
\bra{\psi} &= (\hat{C}\hat{P}T\otimes K\ket{\psi})^{T}
\\&= \sum_{n,m=1}^{2,2} C_{nm}^{*}(\hat{C}\hat{P}T\ket{u_{n}})^{T}\otimes(\ket{u_{m}})^{\dagger}
\end{split}
\end{equation}
where $K=T$ mimics the usual complex conjugation.
\par
The initial density matrix of the composite state \eqref{eq:entstate}, using notations furnished in \eqref{eq:denterm}, is the tensor product
\begin{equation}
\label{eq:denoverall}
\begin{split}
\rho_{1,2} = &\frac{1}{2\cos{\phi}}\begin{bmatrix}
N_{11}& N_{12}\\N_{21}&N_{22}
\end{bmatrix} \otimes \\&\frac{1}{2}\begin{bmatrix}
1+\beta+\beta^{*}& (\delta-\alpha)+(\beta-\beta^{*})\\(\delta-\alpha)-(\beta-\beta^{*})&1-(\beta+\beta^{*})
\end{bmatrix}.
\\
\end{split}
\end{equation}
We immediately infer from \eqref{eq:eigPT} and \eqref{PTQMQM} that finding the partial trace of $\rho_{1,2}$ in either of the Hilbert spaces would retain the same set of eigenvalues. In fact,
if we denote the density matrices by the notations $\rho_{P}(t)$ and $\rho_{S}(t)$ and have the subsystems evolve by means of the operators $U_{1}(t)\otimes\mathbb{I}$ and $\mathbb{I}\otimes U_{2}(t)$ respectively, where $U_{1}(t) = e^{-i\hat{H}_{1}t}$ and $U_{2}(t) = e^{-i\sigma_{x}t}$, then it transpires that for $\rho_{P}(t)$ we have
\begin{equation}
\begin{split}
&\ket{\psi_{t}} =  U_{1}(t)\otimes\mathbb{I}\ket{\psi} = \sum_{n,m=1}^{2,2} e^{-i\lambda_{n}t}C_{nm}\ket{u_{n}}\otimes\ket{v_{m}},
\\ & \lambda_{1} = \sqrt{\zeta^2-\gamma^2},\;\;\;\lambda_{2} = -\sqrt{\zeta^2-\gamma^2},
\\&\rho_{P}(t) = \sum_{n,m,a,b=1}^{2}e^{i(\lambda_{n}-\lambda_{a})t} C_{ab}C_{nm}^{*}\ket{u_{a}}\bra{u_{n}}_{CPT}\otimes\ket{v_{b}}\bra{v_{m}}   
\end{split}
\end{equation}
while for $\rho_{S}(t)$ the following holds
\begin{equation}
\begin{split}
&\ket{\psi_{t}} = \mathbb{I}\otimes U_{2}(t)\ket{\psi} = \sum_{n,m=1}^{2,2} e^{-i\lambda_{m}t}C_{nm}\ket{u_{n}}\otimes\ket{v_{m}},
\\&\lambda_{1}=-1, \;\;\;\lambda_{2} = 1,
\\&\rho_{S}(t) = \sum_{n,m,a,b=1}^{2}e^{i(\lambda_{m}-\lambda_{b})t} C_{ab}C_{nm}^{*}\ket{u_{a}}\bra{u_{n}}_{CPT}\otimes\ket{v_{b}}\bra{v_{m}}.  
\end{split}
\end{equation}
This implies that the entangled state \eqref{eq:entstate} reflects no change in either of the subsystems demonstrating successfully the no-signaling hypothesis. 

\section{Concluding remarks}
The problem of preservation of no-signaling principle is addressed for certain combinations of $PT$-symmetric systems. Since all $PT$-symmetric Hamiltonians are known to belong to the class of pseudo-Hermitian theory, we use the techniques of the latter to establish the result in the affirmative. In this regard, we considered the pair of subsystems governed each by $PTQM$ as one possibility along with $PTQM$ and $SQM$ as another. The key ingredient that we employed is the notion of $CPT$-inner product, which is known to admit of a probabilistic interpretation for a $PTQM$ system, to establish the invariance of the relevant reduced density matrix before and after the operation of time evolution operator. Although the results obtained in this work is deemed to be similar with\cite{Jap2} , we highlight the difference from our work with the elaborate use of modified density matrices, a crucial feature missing in the former.

\section{Acknowledgments}
We thank our colleagues for making several constructive criticisms that, in our opinion, led to a substantial improvement of the paper.

\end{document}